\documentclass[12pt,preprint]{aastex}

\usepackage[dvips]{color}
\shorttitle{R-T Instability in the K-S Prominence Model}
\shortauthors{Hillier et al.}

\begin{document}
\title{Numerical Simulations of the Magnetic Rayleigh-Taylor Instability in the Kippenhahn-Schl\"{u}ter Prominence Model}

\author{Andrew Hillier, Hiroaki Isobe\altaffilmark{1} and Kazunari Shibata}
\affil{Kwasan and Hida Observatories, Kyoto University}
\email{andrew@kwasan.kyoto-u.ac.jp}
\and

\author{Thomas Berger} 
\affil{Lockheed Martin Solar and Astrophysics Laboratory, Palo Alto, CA 94304}

\altaffiltext{1}{Unit of Synergetic Studies for Space, Kyoto University}

\begin{abstract}
The launch of the Hinode satellite has allowed unprecedented high resolution, stable images of solar quiescent prominences to be taken over extended periods of time. 
These new images led to the discovery of dark upflows that propagated from the base of prominences developing highly turbulent profiles. 
As yet, how these flows are driven is not fully understood.
To study the physics behind this phenomena we use 3-D magnetohydrodynamic (MHD) simulations to investigate the nonlinear stability of the Kippenhahn-Shl\"{u}ter prominence model to the magnetic Rayleigh-Taylor instability. 
The model simulates the rise of a buoyant tube inside a quiescent prominence, where the upper boundary between the tube and prominence model is perturbed to excite the interchange of magnetic field lines. 
We found upflows of constant velocity (maximum found $6$\,km\,s$^{-1}$) and a maximum plume width $\approx 1500$\,km which propagate through a height of approximately 6\,Mm in the no guide field case.
The case with the strong guide field (initially $B_y=2B_x$) results in a large plume that rises though the prominence model at $\sim 5$\,km\,s$^{-1}$ with width $\sim 900$\,km (resulting in width of $2400$\,km when viewed along the axis of the prominence) reaching a height of $\sim 3.1$\,Mm.
In both cases nonlinear processes were important for determining plume dynamics.

\end{abstract}

\section{Introduction}\label{intro}

From supernova outflows \citep{HAC1989} to the Earth's ionosphere \citep{Taka2009} the Rayleigh-Taylor instability \citep{TAY1950}, in both its hydrodynamical (HD) and magnetohydrodynamical (MHD) forms, is a fundamental process in astrophysics. 
In Solar physics, the magnetic Rayleigh-Taylor instability is know to be an important driver behind many observed phenomena.
For example the flux emergence process, where magnetic flux rises from the solar interior to the surface and undergoes the magnetic Rayleigh-Taylor instability as it emerges into the upper atmosphere producing observed filamentary structure \citep{IS2006}.

The physical properties of this instability have been well studied.
The growth rate ($\omega$) of the magnetic Rayleigh-Taylor instability for a uniform magnetic field parallel to the interface is:
\begin{equation}
\omega^2=kg\left(A- \frac{B^2 k_{\|}}{4 \pi (\rho_+ + \rho _-) kg} \right )
\end{equation}
where B is the magnetic field strength and $k_{\|}$ is the component perturbation in the direction of the magnetic field and A is the Atwood number, where $A=(\rho_+ - \rho_-)/(\rho_+ + \rho_-) $ with $\rho_+$ the density of the heavier overlying plasma and $\rho_-$ the density of the light plasma underneath  \citep{CHAN1961}.
In the case where $A \rightarrow 1$ the formation of rising bubbles and falling spikes is common \citep{DALY1967}.
\cite{Stone2007} investigated the impact of shear in the magnetic field across the contact discontinuity, finding that this suppressed the small wavenumbers creating wider filamentary structures.  
Recent observations by \cite{BERG2010} of dark upflows that propagate from underdense bubbles through quiescent prominences appear to be the observational signature of the Rayleigh-Taylor instability in quiescent prominences.

Quiescent prominences are large structures of relatively cool \citep[10000 K][]{TH1995}), dense \citep[$\sim 10^{11}$ cm$^{-3}$][]{HIR1986} plasma, that exist in quiet regions of the solar corona, predominantly at high heliographic latitudes. 
Using a characteristic gas pressure of $0.6~dyn~cm^{-2}$ \citep{HIR1986} and magnetic field of $3 \sim 30$ G \citep{LER1989}, gives a plasma $\beta \sim 0.01$-$1$.
Linear magnetohydrostatic modelling of a quiet region filament has shown plasma $\beta \leq 1$ and strong departures from force-free magnetic field \citep{DUD2008}.
Globally , quiescent prominences are incredibly stable structures that often exist in the corona for weeks.
In contrast to this global stability, locally quiescent prominences are highly dynamic phenomena.
Observations of quiescent prominences have shown downflows \citep{ENG1981}, vortices of approximately $10^5$ km $\times 10^5$ km in size \citep{LZ1984} and a bubble of size $2800$ km forming a keyhole shape with a bright center \citep{DT2008} with velocities of 10-30\,km s$^{-1}$.
There are many reviews that give a full description of the current understanding of the structure and dynamics of quiescent prominences \citep[see, for example,][]{TH1995,LAB2010,MAC2010}.

Observations by the Solar Optical Telescope \citep{KOS2007} on the Hinode satellite \citep{TSU2007} have shown that on a small scale quiescent prominences are highly dynamic and unstable phenomena.
\cite{BERG2008} and \cite{BERG2010} reported dark plumes that propagated from large bubbles (approximately $10$\,Mm in size) that form at the base of some quiescent prominences. 
Plumes form at the bubble prominence boundary and then flow through a height of approximately 10\,Mm before dispersing into the background prominence material (see Figure \ref{obs_bubble}).
Observations imply that the plumes and the cavities have a column density about $20$\% of the prominence density \citep{HEINZEL2008}.
The dark upflows maintained an almost constant velocity of approximately $20$\,km\,s$^{-1}$ throughout their rise phase. 
Often these plumes would separate from the large scale bubble forming smaller bubbles inside the prominence material.
\cite{BERG2011} presents observations of large scale prominence bubbles using the Atmospheric Imaging Assembly (AIA) on the Solar Dynamics Observatory (SDO) that show the temperature of the material inside the bubble to be $>250,000$\,$K$.

\cite{BERG2010} hypothesized that the observed upflows were caused by the Rayleigh-Taylor instability.
The observed dynamics seem to be those of a mixed mode perturbation with an interchange mode where a component of $k$, the perturbation wave vector, is perpendicular to the magnetic field $B$ and an undular mode where a component of $k$ is parallel to $B$, in the high Atwood number limit.
\citet{RYU2010} described how the theoretically predicted growth rates and behavior for the magnetic Rayleigh-Taylor instability well match the observations of quiescent prominence plumes.
\citet{VB2010} speculated that the occurrence of the Rayleigh-Taylor instability inside a quiescent prominence could lead to the formation of the thread like structure observed through creation of a tangled magnetic field.
If these observations are created by the Rayleigh-Taylor instability, it has very important implications for our understanding of quiescent prominences as it implies that gravity will play a very important role in the evolution of the system.

\begin{figure}[ht]
\centering
\includegraphics[height=8.3cm]{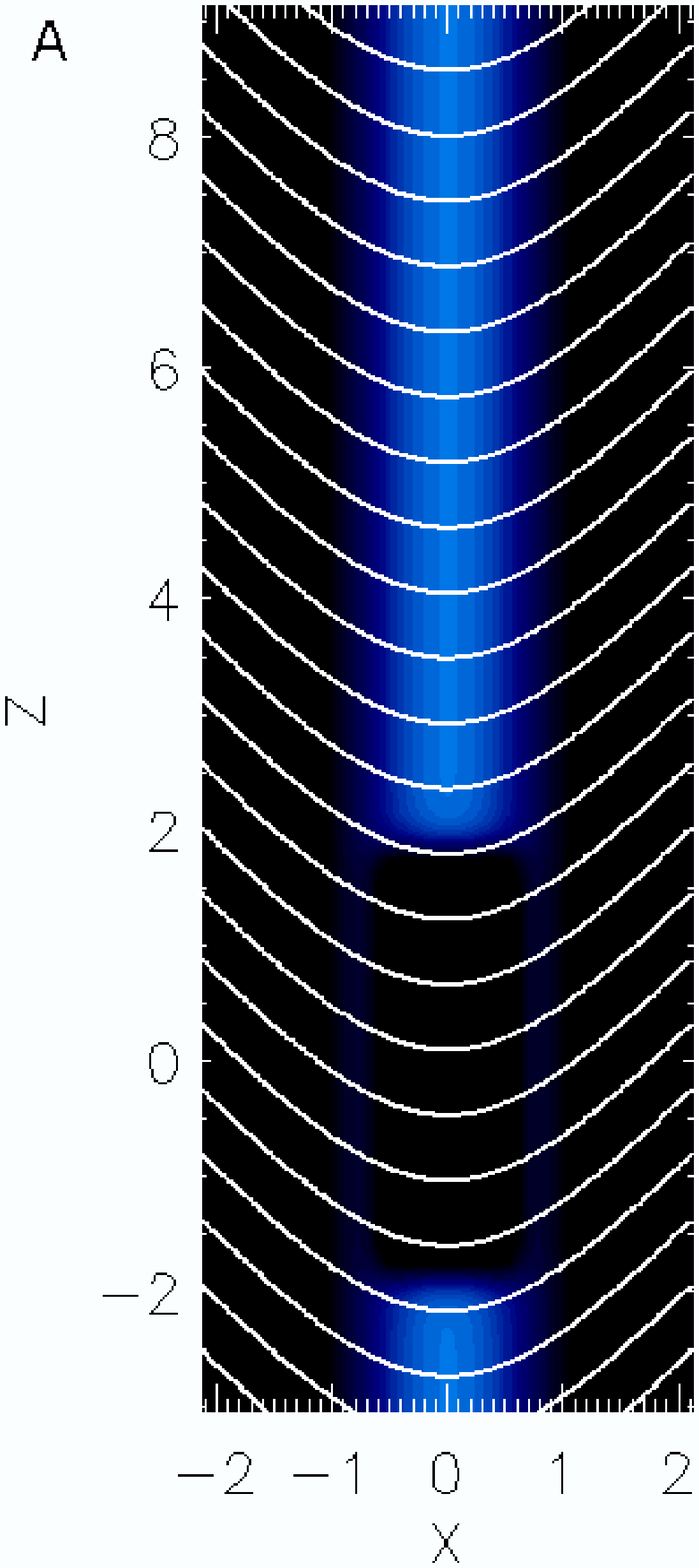}
\includegraphics[height=8.3cm]{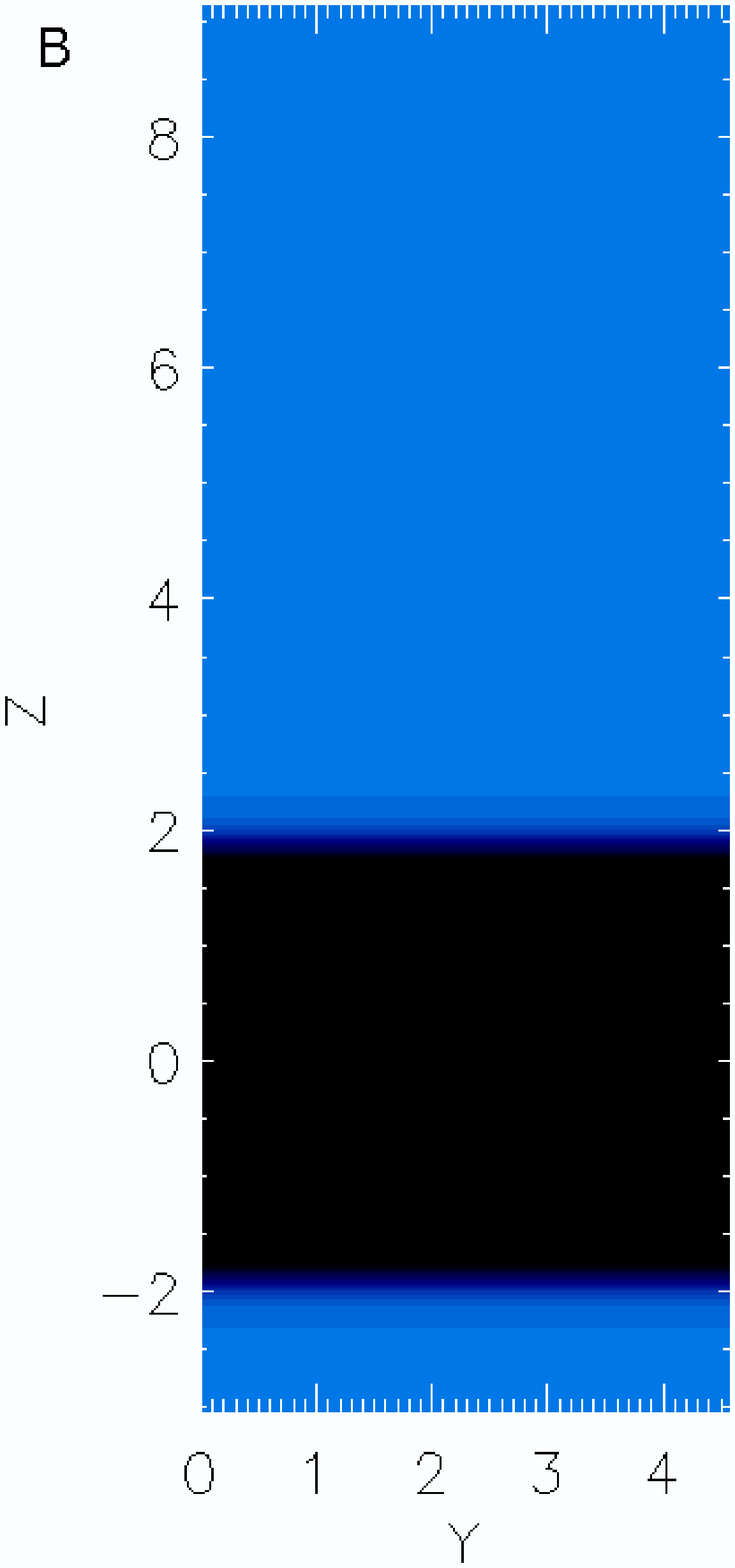}
\includegraphics[height=8.3cm]{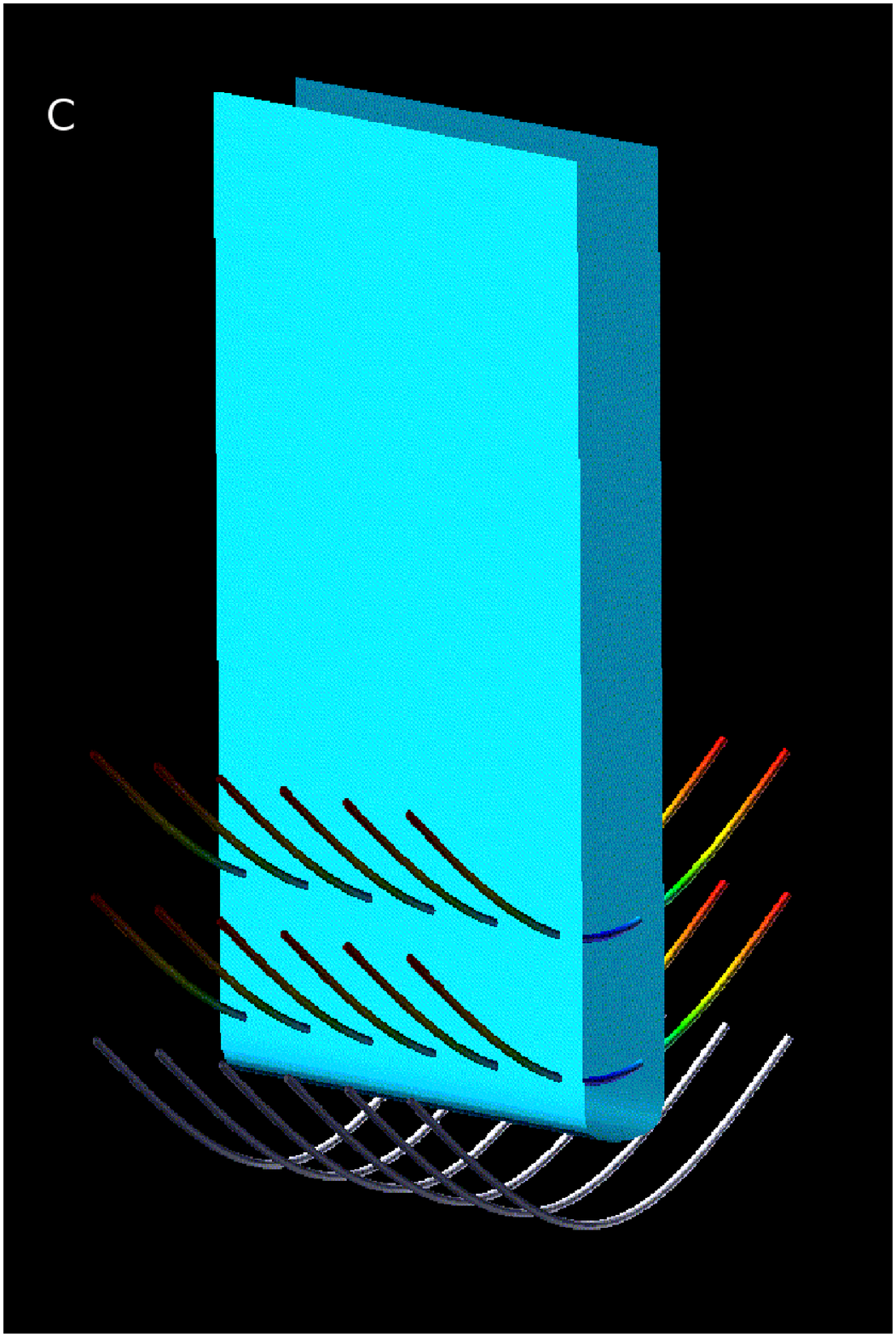}
\caption{Contour plots of the initial density distribution for the standard model for A) the $x-z$ plane at $y=0$ (with magnetic field lines), B) the $y-z$ plane at $x=0$ \& C) the 3D visualisation. All physical quantities are initially constant in the $y$ direction. The initial velocity perturbation is applied to the upper contact discontinuity in along the y direction. The lengths displayed are given in megameters.}
\label{init}
\end{figure}

The model that we use in this work is the Kippenhahn-Schl\"{u}ter model \citep{KS1957, PR1982}).
The Kippenhahn-Schl\"{u}ter model for solar prominences uses the Lorentz force of a curved magnetic field to support plasma against gravity (see Figure \ref{init}) where magnetic tension balances gravity and gas pressure balances magnetic pressure. 
This model describes the local structure of the prominence, without including a corona, and is uniform in the vertical direction.
This model has been shown to be linearly stable to ideal MHD perturbations \citep{KS1957, AN1969}. 

Though this model only describes the local prominence geometry, neglecting the global structure, it can be used to proved an important first step toward modeling these upflows in a prominence like magnetic geometry to understand how the structure of the magnetic field evolves through time, as this can lead to a greater understanding of the quiescent prominence/filament system.
In this paper we present a study of how a nonlinear perturbation in the form of a low density tube placed inside the Kippenhahn-Schl\"{u}ter model can be perturbed to allow the interchange of magnetic field lines causing upflows to form inside the prominence model.

\section{Numerical Method}\label{method}

In this study, we use the 3D conservative ideal MHD equations. 
Constant gravitational acceleration is assumed, but viscosity, diffusion, heat conduction and radiative cooling terms are neglected.
Though the partial ionisation of quiescent prominences may have some effect on the magnetic geometry \citep{HILL2010}, as the timescales of interest here are two orders of magnitude smaller than those associated with the effects of partial ionisation so they have been neglected in this paper. 
We assume the medium to be an ideal gas.
The equations are non-dimentionalized using the sound speed ($C_s=13.2$\,km\,s$^{-1}$), the pressure scale height ($\Lambda=C_s/(\gamma g)= R_gT/(\mu g)=6.1 \times 10^7$ cm), the density at the centre of the prominence ($\rho (x=0)=10^{-13}$ g cm$^{-3})$ giving a characteristic timescale of $\tau=\Lambda/C_s=47$\,s.
The value for $\gamma=1.05$ is taken, this is for simplicity as the internal energy $\epsilon=p/(\gamma-1)$ excludes the possibility of using $\gamma=1$.
Using $\beta=0.5$, which gives an Alfv\'{e}n velocity of $V_A=C_s\sqrt{2/\gamma \beta}=25.8$\,km\,s$^{-1}$.

The initial model is as follows:
\begin{eqnarray}
B_x(x) & = & B_{x0}\\
B_z(x) & = & B_{z\infty} tanh \left(\frac{B_{z\infty}}{2 B_{x0}}\frac{x}{\Lambda} \right)\\
p(x) & = & \frac{B_{z\infty}^2}{8\pi} cosh^{-2} \left(\frac{B_{z\infty}}{2 B_{x0}}\frac{x}{\Lambda} \right)\\
\rho (x) & = &\frac{B_{z\infty}^2}{8\pi} cosh ^{-2} \left(\frac{B_{z\infty}}{2 B_{x0}}\frac{x}{\Lambda} \right) 
\end{eqnarray}
where $B_{x0}$ is the value of the horizontal field at $x=0$ and $B_{z\infty}$ is the value of the vertical field as $x \rightarrow \infty $.
This model is the Kippenhahn-Schl\"{u}ter model as presented in \cite{PR1982}. 

As the Kippenhahn-Schl\"{u}ter model is linearly stable to ideal MHD perturbations, a nonlinear perturbation is necessary.
The perturbation considered here is a high temperature, low density tube placed in the center of the Kippenhahn-Schl\"{u}ter model.
The density of the bubble at $x=z=0$ is $0.3\rho(0)$ (Atwood number $A=0.53$) with width $2\Lambda$ and height $8\Lambda$.
Figure \ref{init} shows a visual representation of the initial conditions.
The color contour represents the mass density, the lines represent the magnetic field lines.
The initially the physical quantities are uniform in the $y$ direction.
To excite the interchange mode a velocity perturbation in $v_y$ was imposed, where $v_y$ was given as a sum of sinusoidal curves of different wavelength.
The maximum amplitude of the perturbation ($|v_y|$) is less than $0.01 C_s$.

To reduce computational time, we assume a reflective symmetry boundary at $x=0$.
Due to the nature of the magnetic field at the top and bottom ($z$) boundary and at $x=L_x$, the choice for boundary is very limited.
A free boundary is assumed at $x=L_x$ with a damping zone (damping time $\tau=4.4$) for the hydromagnetic variables and $B_z$ (to maintain the angle of the magnetic field at the boundary).
This condition is important to maintain tension at the boundary to make sure the plasma can be supported there, but as this is a local simulation this will ultimately have an influence the results.
For the top and bottom boundary, a periodic boundary is assumed and a reflective symmetry boundary is imposed at $y=0,L_y$.

The scheme used is a two step Lax-Wendroff scheme based on the scheme presented in \cite{UGAI08}, using the artificial viscosity and smoothing also presented in this paper. 
The grid size is uniform in the y-direction, and in the x-z plane we take a grid of $75 \times 400$ grid points , where the total area of the calculation domain is $3.5 \Lambda \times 85\Lambda$.
A fine mesh is assigned to an area of $40 \times 320$ grid points, actual size $1.2 \Lambda \times 30\Lambda$, around the upper contact discontinuity allow the plumes to be resolved.
In the $y$-direction a total of $150$ grid points were used with $dy=0.05$.

\section{Evolution of the Upflows}\label{result_RT}

\begin{figure*}[ht]
\includegraphics[width=16.0cm]{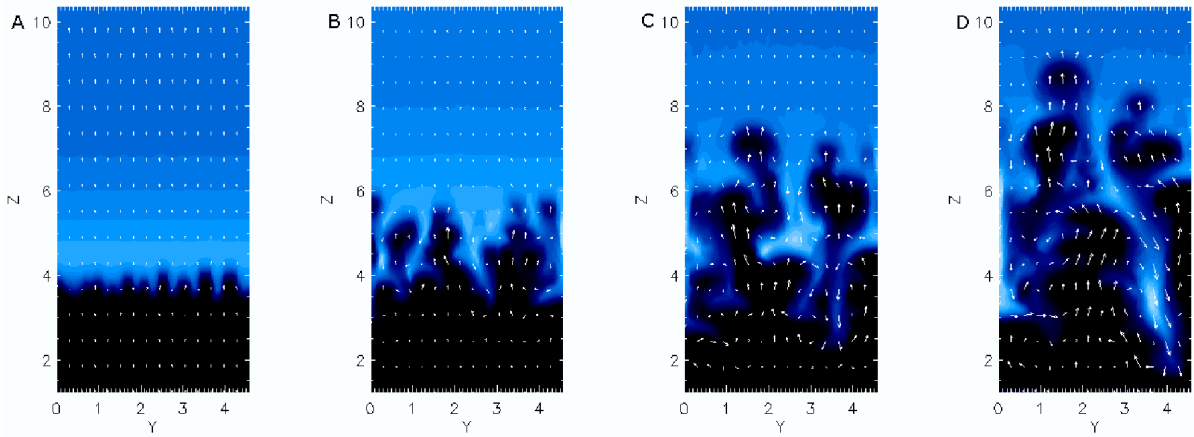}
\caption{Temporal evolution of upflows for $t=719,~1452,~2049$ \& $2453$\,s (normalized units $t=15.3,~30.9,~43.6$ \& $52.2$) taken in the $y - z$ plane at $x=0$. All lengths are given in megameters.}
\label{random}
\end{figure*}

Figure \ref{random} shows the evolution of the upflows for the simulation presented in this paper (where $B_{x0}=B_z{\infty}$ so $\beta=P_{gas}/P_{mag}=0.5)$.
Upflows of size $2\Lambda $-$ 3\Lambda$ in width with velocities $\sim 0.39 C_s$  can be clearly seen in panel D of the figure.

The evolution of the upflows can be understood in the following way.
First the buoyant tube rises inside the prominence as described in the previous section.
Then as the rise of the tube halts the interchange of magnetic field lines is excited by the small perturbation given to the system at the start of the simulation.
The upflows excited have a wavelength of $< 0.5 \Lambda$.
This most unstable wavelength is decided by the numerical viscosity of the scheme and the terms added to the scheme for stability.

As the upflows grow they interact with each other to create larger plumes.
This interaction is driven by the slight difference in plume size created by the random perturbation giving vortexes of different strength and orientation on either side of each plume.
The interaction of the plumes drives stronger vortex motion which creates greater interaction between the plumes.
The result of this interaction is the formation of the large plumes that dominate the system.
This is known as the inverse cascade process and is a common feature of the Rayleigh-Taylor instability (see, for example, \cite{YOUNGS1984} or \cite{IS2006}).

Panels C \& D of figure \ref{random} show the rising bubble and falling spike formation associated with the high Atwood number Rayleigh-Taylor instability.
Though the Atwood number in this case is only $0.53$, due to the significant density difference, these dynamics are observed.
It can be expected that they will become more pronounced as the Atwood number is increased.
It should also be noted that the density difference and magnetic field suppress the role of the Kevin-Helmhotz instability.

After the initial acceleration, the individual upflows all achieve a relatively constant upflow velocity.
This implies that at the top of the plume a force balance is maintained. 
This happens as the initial reduction in tension force from the upflows is balanced by the accumulation of magnetic field at the head of the plume.

\begin{figure}[ht]
\centering
\includegraphics[height=8cm]{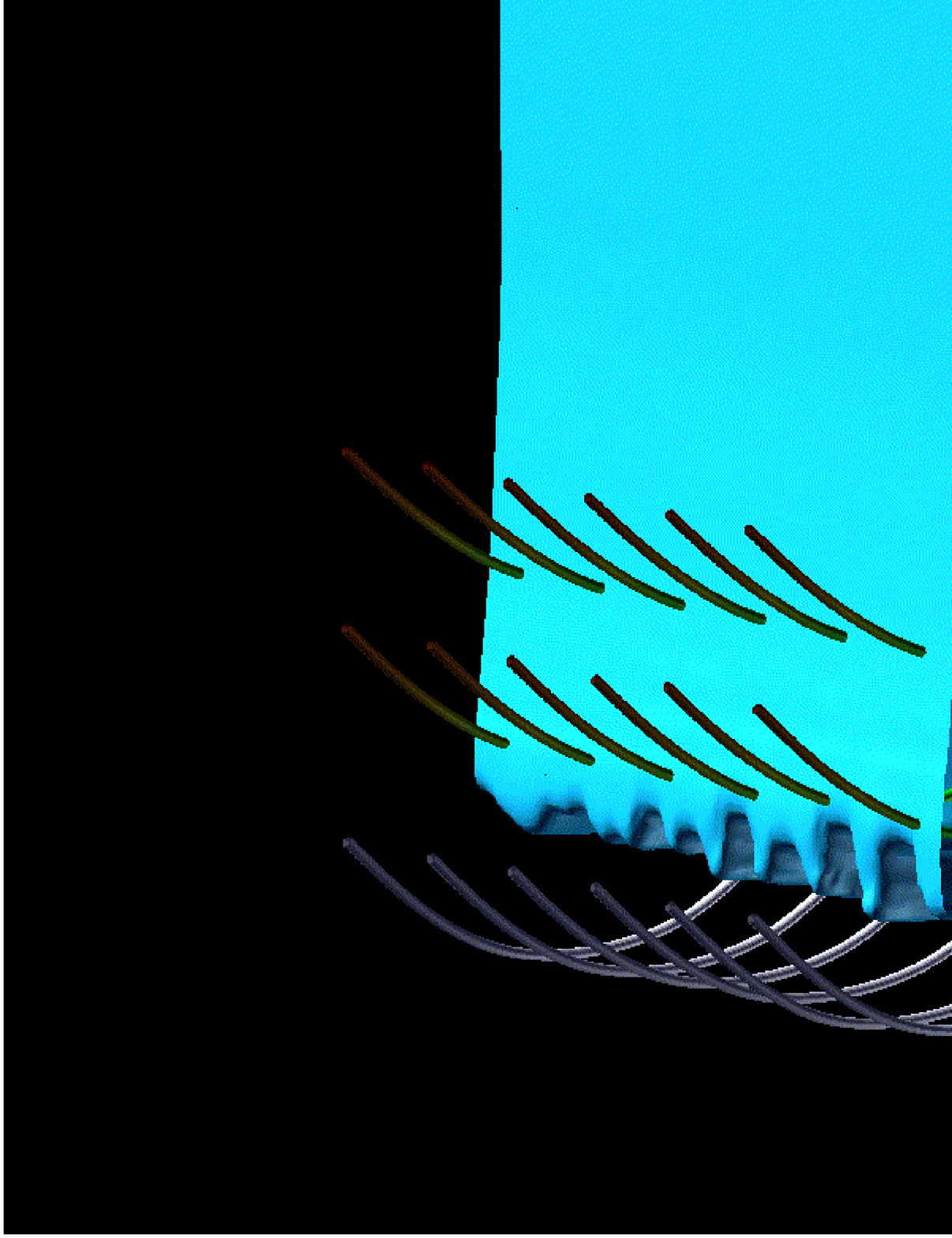}
\includegraphics[height=8cm]{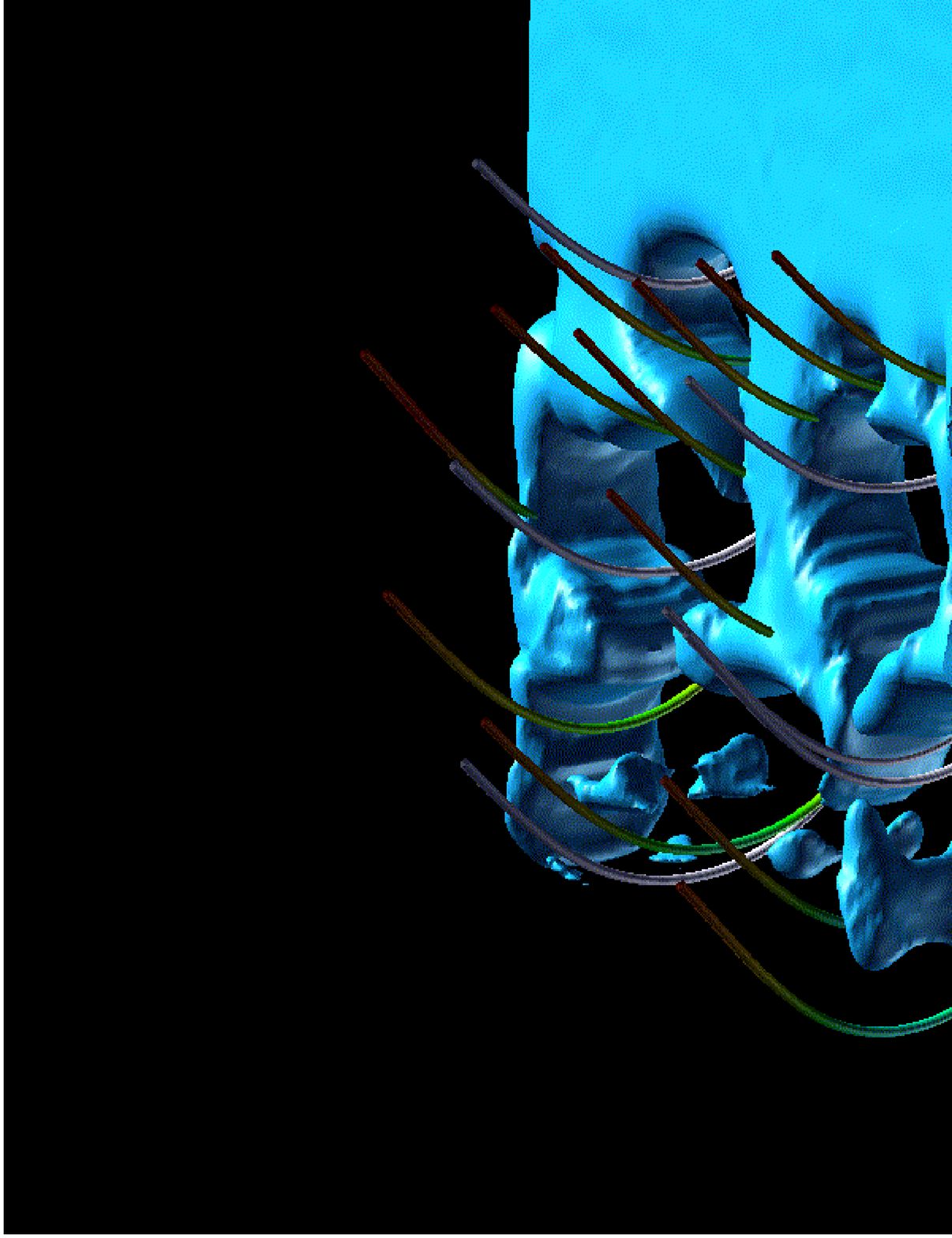}
\caption{3d structure of bubbles created in prominence with seleceted magnetic field lines at times $t=719$ \& $2049$\,s (normalized units $t=15.3$ \& $43.6$.) The online color version distinguishes the field lines that initially thread the hot tube (gray) and those that don't (rainbow) by color.}
\label{3D}
\end{figure}

The 3D structure of the magnetic field evolution caused by the upflows and downflows is displayed in Figure \ref{3D}.
The figure shows the density isosurface at $\rho=0.85$ and the magnetic field lines at $t=15.3$ \& $43.6$. 
Compared to the initial conditions, it is clear that the initial rise of the cavity reduces the tension in the magnetic field around the contact discontinuity.
The figure shows that the upflows form tubes inside the prominence and that the changes in the magnetic field distribution are small.
The rise and fall of the field lines is initiated.
The field lines move by gliding past each other in an interchange process.
As the system evolves, the curvature of the magnetic field lines remains approximately constant after the distortion caused by the initial rise phase.

From this the upflows created inside the Kippenhahn-Schl\"{u}ter model can advect the magnetic field further than the difference between the height of a magnetic field line at $x=0$ and $x=x_{MAX}$.
This process can be seen as analogous to the interchange mode of the magnetic Rayleigh-Taylor instability where the position of magnetic field lines swapped without giving any change to the direction of the magnetic field.

\begin{figure*}[ht]
\centering
\includegraphics[width=16.0cm]{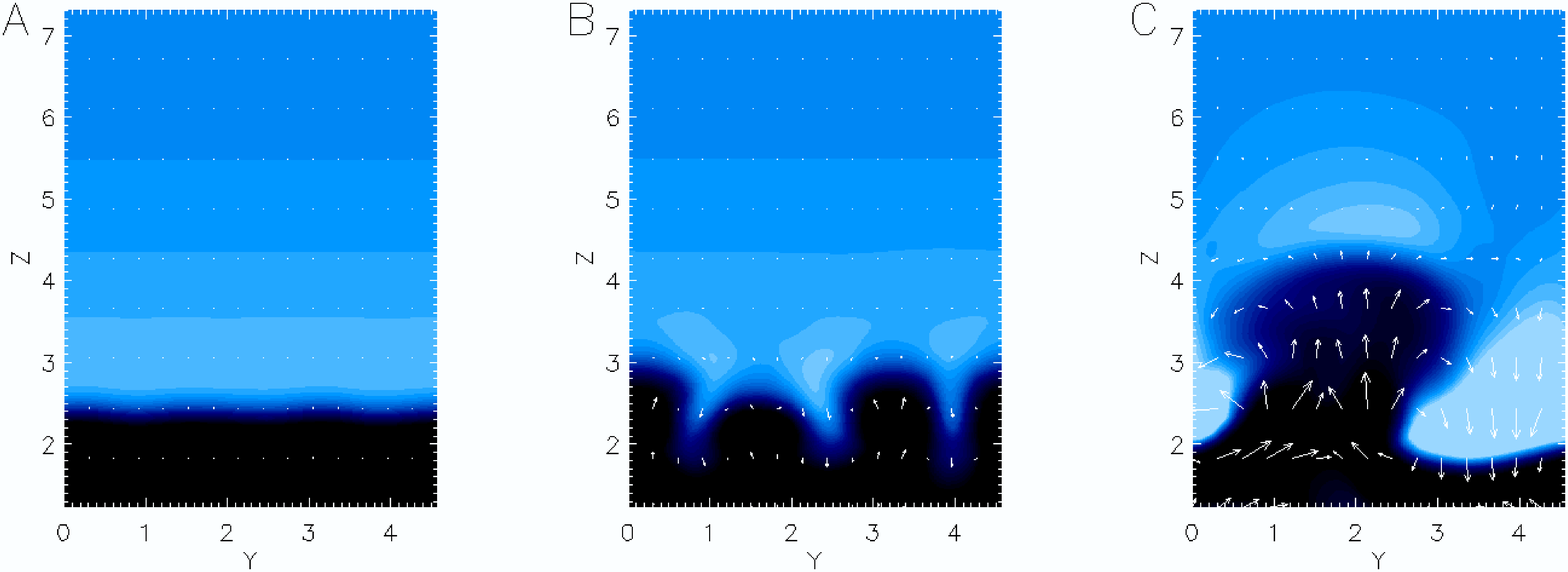}\\
\includegraphics[height=8cm]{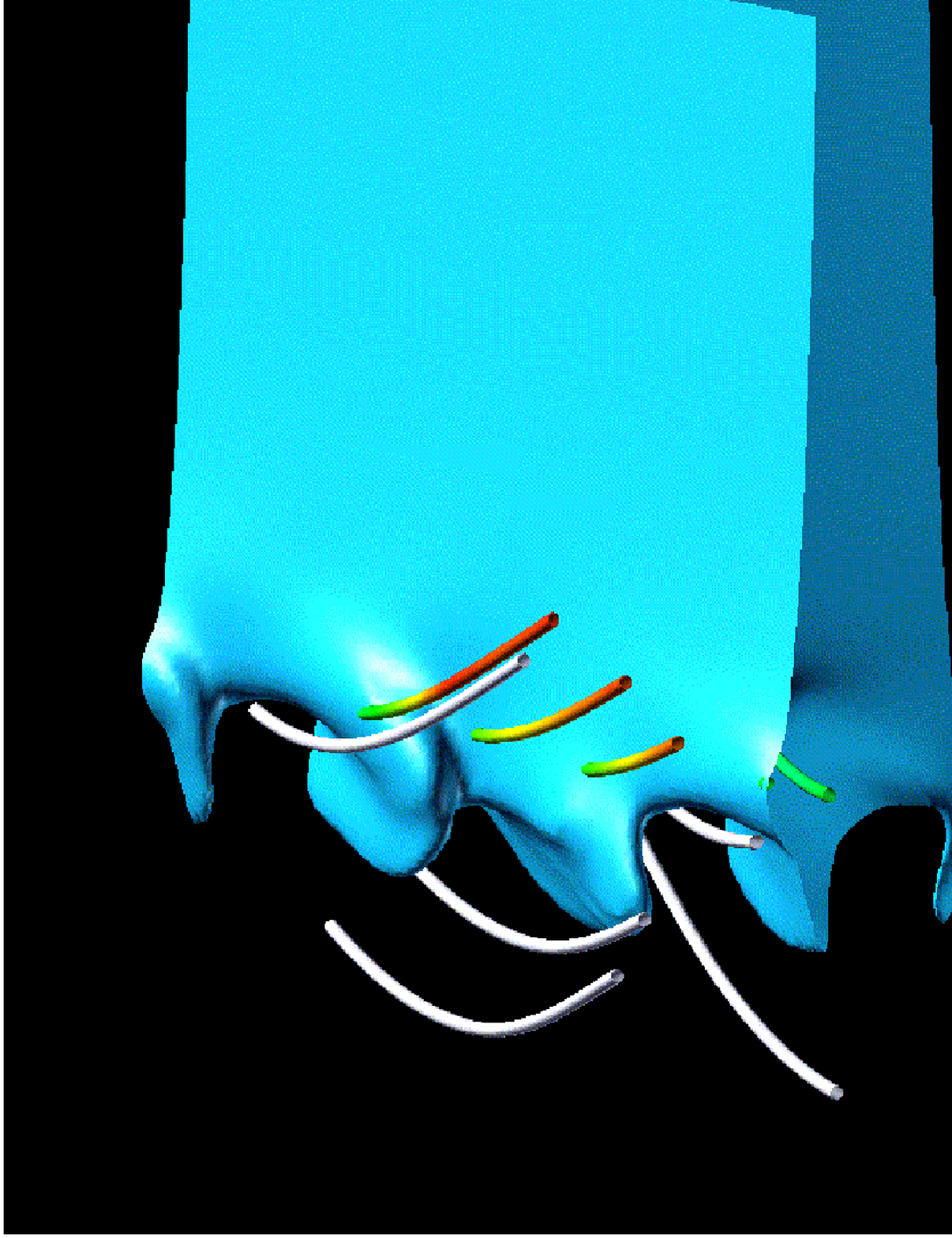}
\includegraphics[height=8cm]{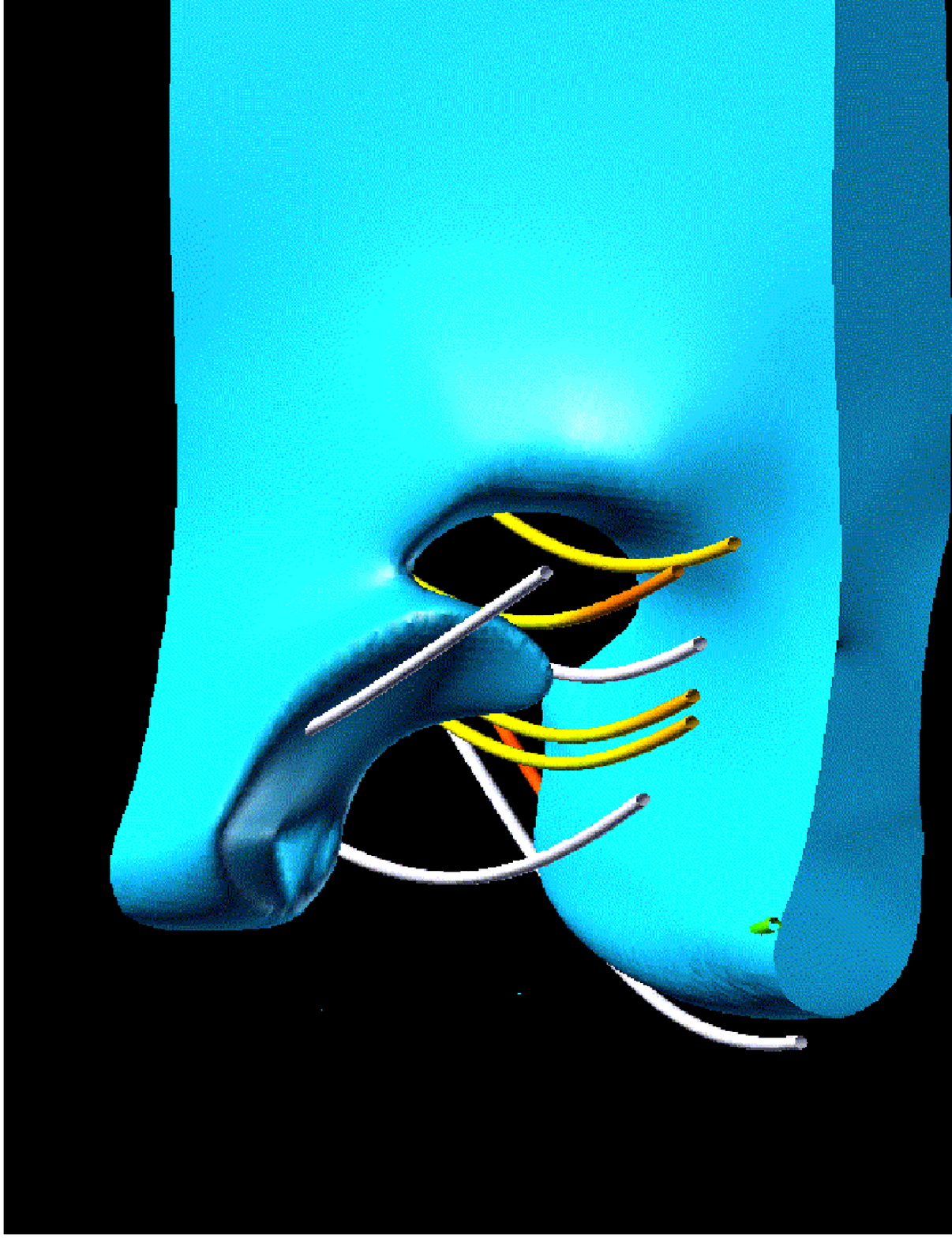}
\caption{Panels A,B \& C show the temporal evolution of upflows for $t=1023,~1646$ \& $2435$\,s (normalized units $t=22.7,~36.6$ \& $52.1$) taken in the $y - z$ plane at $x=0$ for the case with a guide field of strength $B_y=2B_{x0}$. All lengths are given in megameters. Panels D \& E show the 3D structure at times $t=1646$ \& $2435$\,s.}
\label{guide}
\end{figure*}

Figure \ref{guide} shows the evolution of the instability for the same initial conditions as above, but with an initial guide field of $B_y=2B_{x0}$ giving a plasma $\beta=1/6$.
Due to problems with numerical reconnection in current sheets inside that are created in the underdense tube by shear to the guide field, initially the system is allowed to rise in 2D ($x$\,-\,$z$ plane) without a guide field, and once the area around the contact discontinuity has  reached an approximate equilibrium this is taken as the initial conditions for the simualtion.
The guide field is then added and perturbations to this system are solved.

The upper 3 panels of figure \ref{guide} show the 2D slice in the $x=0$ plane.
The initial formation of three plumes of $\sim 1.3$\,Mm in size which through nonlinear coupling produce flows of larger scale.
The instability takes longer to grow than previously and the turbulent flow of the first case is not present.
The 3D images show that the large plumes seen in the 2D images are a result of cutting the filamentary structure at an angle, with the filaments aligned with the magnetic field.
The rise velocity reaches $0.37$\,$C_s$ for a plume of width $\sim 2.4$\,Mm in the $y$\,-\,$z$ plane ($\sim 0.9$\,Mm in the direction perpendicular to the magentic field) with height $\sim 3.1$\,Mm.

\section{Discussion}

In this paper, results of the nonlinear evolution of the magnetic Rayleigh-Taylor instability in the Kippenhahn-Schl\"{u}ter model are presented.
The results show that the nonlinear effects are very important in determining the dynamics of the upflows.
We found that nonlinear mode coupling was important for inducing an inverse cascade forming large upflows from smaller upflows.
The dynamic pressure of the plumes was shown to be able to deform the magnetic field to some extent, but more importantly it allowed the interchange of magnetic field lines.
Therefore, even though the magnetic field did not undergo severe deformation, upflows were be excited.

\begin{figure*}[ht]
\centering
\includegraphics[height=6cm]{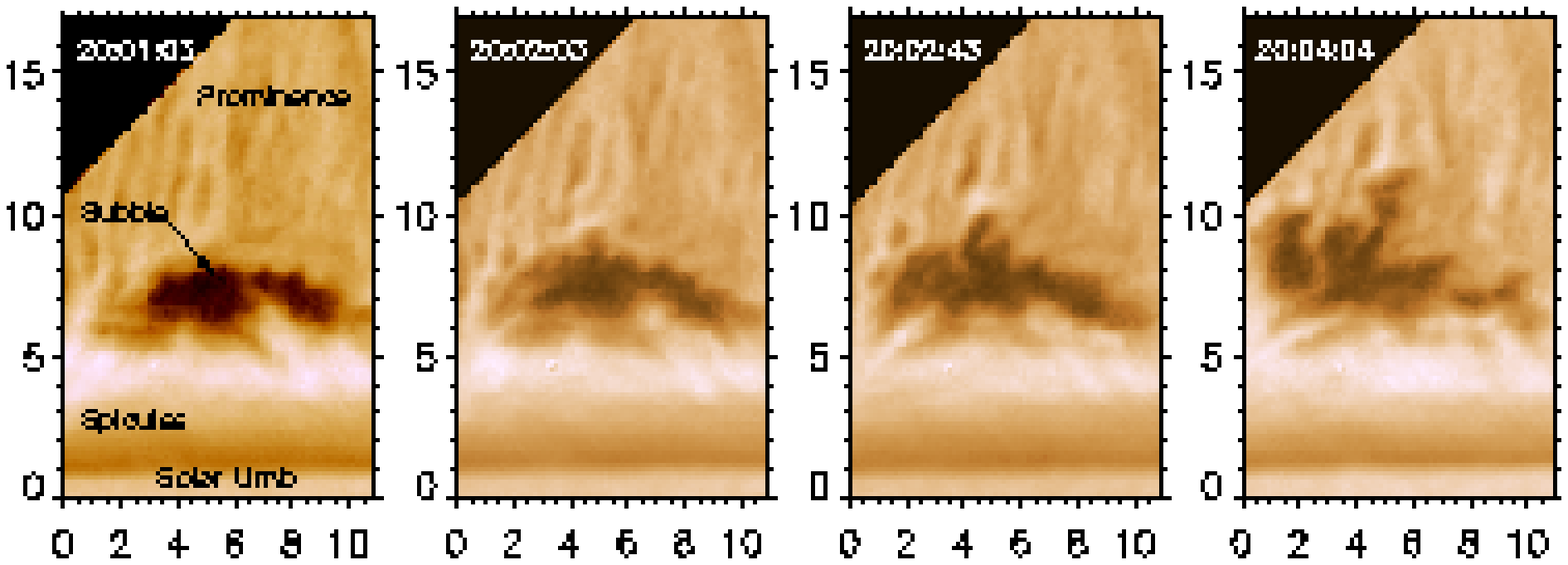}
\caption{Observations showing the formation of dark plumes in a quiescent prominence observed at 8-Aug-2007 20:01 UT observed in the H$\alpha$ line. The axes are labelled in megameters.}
\label{obs_bubble}
\end{figure*}

The model presented here is a local model of a prominence, where the global prominence structure can only be inferred through the boundary conditions.
As this is a local simulation, it must be stressed that the boundary conditions play a strong role in the determination of the system.
Also, it has been assumed that the magnetic field penetrating the hot tube is initially has the same distribution as the cool dense material.
Even with these assumptions, a qualitative comparison between the prominence observations and the plume structure can be performed.
For the comparison with these simulations and observations, the prominences plumes presented in Figure \ref{obs_bubble} are used.
For easy comparison, we will assume that in the observations the magnetic field is aligned along the line of sight.
 
Initially, the bubble forms beneath the prominence and then becomes unstable in multiple areas.
The perturbations form small-scale upflows, which drive the creation of larger scale upflows of a few megameters in width.
These upflows develop highly turbulent profiles.
These upflows described and the upflows found in the simulations presented are very similar in appearance.

To provide a more quantitative comparison the velocity and characteristic sizes can be used.
The observed upflows were found to supersonic, sub Alfv\'{e}nic flows with almost constant velocity of approximately $20$km\,s$^{-1}$.
The simulations develop flows of an almost constant velocity with average velocity of $5.1$\,km\,s$^{-1}$.
This is more than a factor of 3 smaller than the upflows observed.
The observed plumes propagate through a height of approximately $10$\,Mm and have a characteristic width of $\sim 300$\,km$-2$\,Mm.
The simulations produced upflows that have an initial width of $\sim 200$km, but through nonlinear processes produced upflows of $\sim 600$\,km$-1.8$\,Mm in width and in case 1 at the end of the simulation had propagated through a height of $6$\,Mm.

The inclusion of a guide field does not change the dynamics drastically, where the main differences are that the instability takes longer to grow and creates much larger structures in the $y$\,-\,$z$ plane.
The smaller growth rate could be a result of the stronger field strength suppressing the perturbation.
The increase in field strength is likely to be the cause for the visible reduction in the turbulent like profile, as in this case the magnetic pressure dominates the dynamic and gas pressure.
It will be important to investigate a guide field that creates shear between the magnetic field of the large bubble and the prominence, as the size of the filaments is likely to be dependent on the shear of the magnetic field \citep{Stone2007}.

In this paper, a boundary that keeps the magnetic field pointing upward at the boundary is used.
This boundary confines the prominence material, chosen because of the lack of downflows of prominence material through the large bubbles formed at the base of prominences due to upwardly curved magnetic field. 
A fixed boundary was not used due to the clear motion of plasma (and with it magnetic field).
Therefore the boundary used was seen as the most probable.
To elucidate this further, a full study of the boundary will be presented in a separate paper.

The paper presents the first efforts to simulate the formation of upflows in a solar prominence. 
The observation of these prominence upflows has presented many difficult questions, among which are `how do turbulent flows develop in the low-beta environment of prominences?' and `what is the origin of the upflow plumes and how do their dynamics depend on factors such as bubble density and prominence magnetic field strength?' 
The answers to these questions will have implications not only for prominence science, but for the study of complex flows in any low-beta magnetic plasma environment, for example the Equatorial Spread-F phenomenon in the Earth's ionosphere \citep{Taka2009}.
To fully understand the dynamics of these prominence bubbles and the plumes generated therefrom will require a wide parameter survey using the model developed here.

\bigskip

Hinode is a Japanese mission developed and launched by ISAS/JAXA, with NAOJ as domestic partner and NASA and STFC (UK) as international partners. 
It is operated by these agencies in co-operation with ESA and NSC (Norway).
We would like to thank the staff members of Kwasan and Hida observatories for their support.
TEB was supported by NASA contract NNM07AA01C (Solar-B FPP) at Lockheed Martin.
This work was supported by the Grant-in-Aid for the Global COE program ``The Next Generation of Physics, Spun from Universality and Emergence'' from the Ministry of Education, Culture, Sports, Science and Technology (MEXT) of Japan.
HI is supported by the Grant-in-Aid for Young Scientists (B, 22740121).

\begin{thebibliography}{}
\bibitem[Aulanier \& D{\'e}moulin(2003)]{AUL2003} Aulanier, G., \& D{\'e}moulin, P.\ 2003, \aap, 402, 769 

\bibitem[Anzer (1969)]{AN1969} Anzer, U.\ 1969, \solphys, 8, 37 

\bibitem[Berger et al.(2008)]{BERG2008} Berger, T.~E., et al.\ 2008, \apjl, 676, L89 

\bibitem[Berger at al.(2010)]{BERG2010} Berger, T.~E., et al.\ 2010, \apj, 716, 1288

\bibitem[Berger et al.(2011)]{BERG2011} Berger, T., et al.\ 2011, \nat, 472, 197 

\bibitem[Chandrasekhar(1961)]{CHAN1961} Chandrasekhar, S.\ 1961, International Series of Monographs on Physics, Oxford: Clarendon, 1961,

\bibitem[Daly(1967)]{DALY1967} Daly, B.~J.\ 1967, Physics of Fluids, 10, 297 

\bibitem[de Toma et al.(2008)]{DT2008}de Toma, G., Casini, R., Burkepile, J.~T., \& Low, B.~C.\ 2008, \apjl, 687, L123 

\bibitem[Dud{\'{\i}}k et al.(2008)]{DUD2008} Dud{\'{\i}}k, J., Aulanier, G., Schmieder, B., Bommier, V., \& Roudier, T.\ 2008, \solphys, 248, 29 

\bibitem[Engvold(1981)]{ENG1981} Engvold, O.\ 1981, \solphys, 70, 315 

\bibitem[Hachisu et al.(1992)]{HAC1989} Hachisu, I., Matsuda, T., Nomoto, K., \& Shigeyama, T.\ 1992, \apj, 390, 230 

\bibitem[Heinzel et al.(2008)]{HEINZEL2008} Heinzel, P., et al.\ 2008, \apj, 686, 1383 

\bibitem[Hillier et al.(2010)]{HILL2010} Hillier, A., Shibata, K., \& Isobe, H.\ 2010, \pasj, 62, 1231 

\bibitem[Hirayama(1986)]{HIR1986} Hirayama, T.\ 1986, NASA Conference Publication, 2442, 149

\bibitem[Isobe et al.(2006)]{IS2006} Isobe, H., Miyagoshi, T., Shibata, K., \& Yokoyama, T.\ 2006, \pasj, 58, 423 

\bibitem[Kippenhahn \& Schl\"{u}ter(1957)]{KS1957} Kippenhahn, R., \& Schl{\"u}ter, A.\ 1957, \zap, 43, 36 

\bibitem[Kosugi at al.(2007)]{KOS2007} Kosugi, T., et al.\ 2007, \solphys, 243, 3 

\bibitem[Kubota \& Uesugi(1986)]{KUBO1986} Kubota, J., \& Uesugi, A.\ 1986, \pasj, 38, 903

\bibitem[Labrosse et al.(2010)]{LAB2010} Labrosse, N., Heinzel, P., Vial, J.-C., Kucera, T., Parenti, S., Gun{\'a}r, S., Schmieder, B., \& Kilper, G.\ 2010, \ssr, 151, 243 

\bibitem[Lerche \& Low(1980)]{LELO1980} Lerche, I., \& Low, B.~C.\ 1980, \solphys, 67, 229 

\bibitem[Leroy(1989)]{LER1989} Leroy, J.~L.\ 1989, Dynamics and Structure of Quiescent Solar Prominences, 150, 77 

\bibitem[Liggett and Zirin(1984)]{LZ1984} Liggett, M., \& Zirin, H.\ 1984, \solphys, 91, 259

\bibitem[Mackay et al.(2010)]{MAC2010} Mackay, D.~H., Karpen, J.~T., Ballester, J.~L., Schmieder, B., \& Aulanier, G.\ 2010, \ssr, 151, 333 

\bibitem[Priest(1982)]{PR1982} Priest, E.~R.\ 1982, Dordrecht, Holland ; Boston : D.~Reidel Pub.~Co.~; Hingham,, 74P

\bibitem[Ryutova et al.(2010)]{RYU2010} Ryutova, M., Berger, T., Frank, Z., Tarbell, T., \& Title, A.\ 2010, \solphys, 170

\bibitem[Stone \& Gardiner(2007)]{Stone2007} Stone, J.~M., \& Gardiner, T.\ 2007, \apj, 671, 1726

\bibitem[Takahashi et al.(2009)]{Taka2009} Takahashi, H., et al.\ 2009, Annales Geophysicae, 27, 1477 

\bibitem[Tandberg-Hanssen(1995)]{TH1995} Tandberg-Hanssen, E.\ 1995, Astrophysics and Space Science Library, 199, 

\bibitem[Taylor(1950)]{TAY1950} Taylor, G.\ 1950, Royal Society of London Proceedings Series A, 201, 192 

\bibitem[Tsuneta et al.(2007)]{TSU2007} Tsuneta, S., et al.\ 2008, \solphys, 249, 167

\bibitem[Ugai(2008)]{UGAI08} Ugai, M.\ 2008, Physics of Plasmas, 15, 082306 

\bibitem[van Ballegooijen \& Cranmer(2010)]{VB2010} van Ballegooijen, A.~A., \& Cranmer, S.~R.\ 2010, \apj, 711, 164

\bibitem[Youngs(1984)]{YOUNGS1984} Youngs, D.~L.\ 1984, Physica D Nonlinear Phenomena, 12, 32 


\end{thebibliography}
\end{document}